\documentstyle[12pt]{article}

\begin{document}
\thispagestyle{empty}

\begin{center}
\LARGE \tt \bf{On the non-minimal coupling of Riemann-flat Klein-Gordon
Fields to Space-time torsion}

\end{center}

\vspace{1cm}

\begin{center} {\large L.C. Garcia de Andrade\footnote{Departamento de
F\'{\i}sica Te\'{o}rica - Instituto de F\'{\i}sica - UERJ

Rua S\~{a}o Fco. Xavier 524, Rio de Janeiro, RJ

Maracan\~{a}, CEP:20550-003 , Brasil.

E-Mail.: GARCIA@SYMBCOMP.UERJ.BR}}
\end{center}

\vspace{1.0cm}

\begin{abstract}
The energy spectrum of Klein-Gordon particles is obtained via the 
non-minimal coupling of Klein-Gordon fields to Cartan torsion in 
the approximation of Riemann-flatness and constant torsion.When the 
mass squared is proportional to torsion coupling constant it is shown that the splitting of energy 
does not occur.I consider that only the vector part of torsion does 
not vanish and that it is constant.A torsion Hamiltonian operator is 
constructed.The spectrum of Klein-Gordon fields is continuos.
 
\end{abstract}      
\vspace{1.0cm}       
\begin{center}
\end{center}

\newpage
\pagestyle{myheadings}

\section{Introduction}
\paragraph*{}
Earlier Figueiredo,Dami\~{a}o Soares and Tiomno (FTD) 
\cite{1,2,3} investigated the gravitational coupling of 
Klein-Gordon  and Dirac fields to matter vorticity and 
torsion in some cosmological models.Some years earlier 
H.Rumpf \cite{4} considered a particular case of that 
problem namely a Riemann-flat spectrum for Dirac 
particles in the case of constant torsion and 
electromagnetic fields.More recently Claus 
L\" {a}mmerzhal \cite{5} have considered the case of 
the coupling of space-time torsion to the Dirac equation showing the effects on 
the energy levels of atoms which can be tested by the 
Hughes-Drever experiments.In this paper he was able to 
place a limit on the axial torsion testing the 
anisotropy of anomalous spin couplings and mass.Yet 
more recently I.L.Shapiro \cite{6} and Bagrov, 
Buchbinder and Shapiro \cite{7} considered the 
possibility of testing torsion theories in the low 
energy limit by making use of a non-Minimal coupling 
Lagrangian with torsion and scalar fields.At this point 
is important to note that in references \cite{1,2,3,4,5}
the authors have considered that the torsion did not 
propagate or in other words they considered to be 
dealing with the Einstein-Cartan gravity torsion does 
not propagate.Here on the contrarywe consider the 
general case where torsion propagates and besides we 
assume a non-minimal coupling where the Klein-Gordon 
scalar fields couple with torsion, contrary to the (FDT) 
approach where only Dirac fields couple to torsion 
through the spinorial connection.This is also the 
point of view addopted by  Carroll and Field \cite{8} 
who showed that for a wide class of models the only 
modes of the torsion tensor which interact with matter 
are either a massive scalar or a massive spin-1 boson.
In fact the Lagrangian we addopt here is nothing but a 
slight modification of their Lagrangian
\begin{equation}
L=a{\partial}_{\mu}T_{\nu}{\partial}^{\mu}T^{\nu}+b({\partial}_{\mu}T^{\mu})^{2}+cT_{\mu}T^{\mu}
\label{1}
\end{equation}
where $ T^{\mu} $ is the torsion pseudo-vector.Our 
Lagrangian is given by 
\begin{equation}
L_{T}=L ={\partial}_{\mu}{\phi}{\partial}^{\mu}{\phi}+{\partial}_{\mu}T^{\mu}+T_{\mu}T^{\mu}-m^{2}{\phi}^{2}
\label{2}
\end{equation}
This Lagrangian can in fact be obtained from the 
following Lagrangian
\begin{equation}
L={\partial}_{\mu}{\phi}{\partial}^{\mu}{\phi}+ sR{\phi}^{2}-m^{2}{\phi}^{2}
\label{3}
\end{equation}
Where  $ s $ represents the torsion coupling and the last term is the potential energy term and the other terms in equation (\ref{3}) are obtained by 
considering that the scalar curvature $ R $ is Riemann-
flat and given by 
\begin{equation} 
R={\partial}_{\mu}T^{\mu}+T_{\mu}T^{\mu}
\label{4 }
\end{equation}
To simplify matters we shall from now on suppress 
indices.Thus variation of Lagrangian (\ref{2}) with 
respect to torsion and the scalar field $ {\phi} $ 
yields respectively the equations
\begin{equation}
T=-\frac{{\partial}{\phi}}{\phi}
\label{5}
\end{equation}
and
\begin{equation}
{{\partial}^{2}}{\phi}+s{\partial}{\phi}-\frac{{m}^{2}}{2}{\phi}=0
\label{6}
\end{equation}
Let us now solve this system for the case of plane 
symmetry similar to the case of domain walls in 
Riemann-Cartan space-time \cite{9}).In this case the 
above equations may be written in the form
\begin{equation}
{\phi}^{"}+s{\phi}^{`}-\frac{{m}^{2}}{2}{\phi}=0
\label{7}
\end{equation}
where the upper primes denote derivatives with respect 
to coordinate-z.Note that the second term in equation 
(\ref{7}) represents a damping like a friction in domain
walls,thus one may say that torsion introduces a sort of
friction into the problem.Discrete spectrum can be found
for example in refence \cite{1} in the case of the 
G\"{o}del cosmological model.The LHS of equation (\ref{7}) can be 
written in operator form as
\begin{equation}
\hat{H}{\phi}=(\hat{H}_{0}+\hat{H}_{torsion}){\phi}
\label{8}
\end{equation}
where $ \hat{H}_{0} $ is the basic Klein-Gordon Hamiltonian 
operator and $ \hat{H}_{torsion} $ is the Hamiltonian torsion 
operator given by
\begin{equation}
\hat{H}_{torsion}= s{\frac{\partial}{{\partial}z}}
\label{9}
\end{equation}
A similar operator but much more involved have been 
constructed by L\"{a}mmerzhal \cite{5}.To solve 
(\ref{6}) we make use of a simple ansatz 
\begin{equation}
{\phi}=e^{cz}
\label{10}
\end{equation}
Substitution of (\ref{10}) into (\ref{6}) yields the 
following constraint equation
\begin{equation}
c^{2}+sc+m^{2}=0
\label{11}
\end{equation}
 
which is an algebraic very simple equation which yields
\begin{equation}
c=\frac{-s \pm (s^{2}-4m^{4})^{\frac{1}{2}}}{2}
\label{12}
\end{equation}
substitution of (\ref{12}) into (\ref{10}) yields the 
scalar field
\begin{equation}
{\phi}=e^{\frac{-s \pm (s^{2}-4m^{4})^{\frac{1}{2}}}{2}z}
\label{13}
\end{equation}
one may calculate from (\ref{13}) the following energy
\begin{equation}
{\epsilon}=|{\phi}^{'}|^{2}+V({\phi})
\label{14}
\end{equation}
substitution of (\ref{13}) into (\ref{14}) yields
\begin{equation}
{\epsilon}=|\frac{-s \pm (s^{2}-4m^{4})^{\frac{1}{2}}}{2}|^{2}e^{cz}
\label{15}
\end{equation}
From this last expression one notes immediatly that 
there is a splitting of the spectral lines of the 
Klein-Gordon fields always that torsion does not 
coincide with the double of the mass squared.The 
physical constraint required here unfortunatly makes 
the task of measuring torsion directly since at least 
in astrophysical stellarobjects the mass is always much 
higher the torsion imagine the mass squared.Nevertheless
indirect measurements can be sugested.Although there is 
a double splitting in the energy one must observe that 
this spectrum is continuos.In reference (\ref{1}) a 
similar situation appears, nevertheless in their case 
vorticity plays the role of mass here.In other words is 
the simultaneous presence of torsion and mass here that 
produces the splitting of the energy levels.It is also 
important to note that the torsionless case is not 
allowed here since in this case the energy would be a 
complex number.Discrete spectrum can also be found in   
reference \cite{1} in the case of the G\"{o}del 
cosmological model.Yet in reference \cite{7} Bagrov et al. obtained a double splitting of the spectral lines of the Hydrogen atom by considering the low energy limit of torsion in the Schr\"{o}dinger equation.Also in their case the splitting is a pure torsion effect and does not depend on the magnetic field.A more detailed analysis of these 
experiments may appear elsewhere.Besides a small change in the potential in our Lagrangian would allows to investigate gravitational torsion kinks or domain walls.    
\section*{Acknowledgments}
\paragraph*{}
I am very grateful to Prof.Claus L\"{a}mmerzhal for 
providing fundamental ideas for the developement of 
this work.Thanks are also due to to Prof.Ilya Shapiro 
and my colleagues Jim Skea and Rudnei Ramos for helpful 
discussions on the subject of this paper.Thanks are 
due to CNPq. and DAAD(Bonn) from financial support.

\end{document}